\documentclass[12pt]{article}
\usepackage{amsmath}
\usepackage{amsthm}
\usepackage{amsfonts}
\usepackage[dvips]{epsfig}
\usepackage{graphicx}
\usepackage{amssymb}
\usepackage{cancel}
\usepackage{pstricks} 
\usepackage{lscape}
\usepackage{enumitem}
\usepackage{comment}
\usepackage{subcaption}

\usepackage[sort&compress,numbers, merge]{natbib}

\DeclareFontFamily{U}{mathx}{\hyphenchar\font45}
\DeclareFontShape{U}{mathx}{m}{n}{<-> mathx10}{}
\DeclareSymbolFont{mathx}{U}{mathx}{m}{n}

\newcommand{\beq}{\begin{equation}}
\newcommand{\eeq}{\end{equation}}
\newcommand{\bea}{\begin{eqnarray}}
\newcommand{\eea}{\end{eqnarray}}

\makeatletter
\newlength{\apb@width}
\newcommand{\autoparbox}[2][c]{\settowidth{\apb@width}{#2}\parbox[#1]{\apb@width}{#2}}

\newcommand{\Cen}[2]{%
  \ifmeasuring@
    #2%
  \else
    \makebox[\ifcase\expandafter #1\maxcolumn@widths\fi]{$\displaystyle#2$}%
  \fi
}
\makeatother

\definecolor{Orange}{cmyk}{0,0.61,0.87,0}
\definecolor{JungleGreen}{cmyk}{0.99,0,0.52,0}
\definecolor{OliveGreen}{cmyk}{0.64,0,0.95,0.40}
\definecolor{Brown}{cmyk}{0,0.81,1,0.60}
\definecolor{RoyalBlue}{cmyk}{0.71,0.53,0,0.12}

\allowdisplaybreaks[1]

\usepackage[colorlinks=true, linkcolor=OliveGreen, citecolor=RoyalBlue,
urlcolor=RoyalBlue]{hyperref}

\begin{document}

\def\a{\alpha}
\def\b{\beta}
\def\c{\varepsilon}
\def\d{\delta}
\def\e{\epsilon}
\def\f{\phi}
\def\g{\gamma}
\def\h{\theta}
\def\k{\kappa}
\def\l{\lambda}
\def\m{\mu}
\def\n{\nu}
\def\p{\psi}
\def\q{\partial}
\def\r{\rho}
\def\s{\sigma}
\def\t{\tau}
\def\u{\upsilon}
\def\v{\varphi}
\def\w{\omega}
\def\x{\xi}
\def\y{\eta}
\def\z{\zeta}
\def\D{\Delta}
\def\G{\Gamma}
\def\H{\Theta}
\def\L{\Lambda}
\def\F{\Phi}
\def\P{\Psi}
\def\S{\Sigma}

\def\o{\over}
\def\beq{\begin{eqnarray}}
\def\eeq{\end{eqnarray}}
\newcommand{\gsim}{ \mathop{}_{\textstyle \sim}^{\textstyle >} }
\newcommand{\lsim}{ \mathop{}_{\textstyle \sim}^{\textstyle <} }
\newcommand{\vev}[1]{ \left\langle {#1} \right\rangle }
\newcommand{\bra}[1]{ \langle {#1} | }
\newcommand{\ket}[1]{ | {#1} \rangle }
\newcommand{\EV}{ {\rm eV} }
\newcommand{\KEV}{ {\rm keV} }
\newcommand{\MEV}{ {\rm MeV} }
\newcommand{\GEV}{ {\rm GeV} }
\newcommand{\TEV}{ {\rm TeV} }
\def\diag{\mathop{\rm diag}\nolimits}
\def\Spin{\mathop{\rm Spin}}
\def\SO{\mathop{\rm SO}}
\def\O{\mathop{\rm O}}
\def\SU{\mathop{\rm SU}}
\def\U{\mathop{\rm U}}
\def\Sp{\mathop{\rm Sp}}
\def\SL{\mathop{\rm SL}}
\def\tr{\mathop{\rm tr}}

\def\IJMP{Int.~J.~Mod.~Phys. }
\def\MPL{Mod.~Phys.~Lett. }
\def\NP{Nucl.~Phys. }
\def\PL{Phys.~Lett. }
\def\PR{Phys.~Rev. }
\def\PRL{Phys.~Rev.~Lett. }
\def\PTP{Prog.~Theor.~Phys. }
\def\ZP{Z.~Phys. }

\vspace{0.5cm}
\begin{center}
{\bf \Large W boson mass anomaly and grand unification
}
\end{center}
\vspace{0.75cm}

\begin{center}
{\bf Jason~L.~Evans}$^{a}$,
{\bf Tsutomu~T.~ Yanagida}$^{a,b}$, and
{\bf Norimi Yokozaki}$^c$
\end{center}

\begin{center}
  {\em $^a$Tsung-Dao Lee Institute, Shanghai Jiao Tong University, Shanghai 200240, China}\\[0.2cm] 
  {\em $^b$Kavli IPMU (WPI), UTIAS, University of Tokyo, Kashiwa, Chiba 277-8583, Japan}\\[0.2cm] 
{\em $^c$Zhejiang Institute of Modern Physics and Department of Physics, Zhejiang University, Hangzhou, Zhejiang 310027, China
}\\[0.2cm]

\end{center}

\vspace{1cm}
\centerline{\bf ABSTRACT}
\vspace{0.2cm}

{\small 
It is known that the recently reported shift of the W boson mass can be easily explained by an $SU(2)_L$ triplet Higgs boson with a zero hypercharge if it obtains a vacuum expectation value (VEV) of $O(1)$ GeV. Surprisingly, the addition of a TeV scale complex triplet Higgs boson to the standard model (SM) leads to a precise unification of the gauge couplings at around $10^{14}$\,GeV. We consider that it is a consequence of $SU(5)$ grand unification and show a possible potential for the Higgs fields yielding a weak scale complex $SU(2)$ triplet scalar boson. Although it seems the proton decay constraint would doom such a low-scale unification, we show that the constraint can be avoided by introducing vector-like fermions which mix with the SM fermions through mass terms involving the VEV of GUT breaking Higgs field. Importantly, the simplest viable model only requires the addition of one pair of vector-like fermions transforming ${\bf 10}$ and $\overline{{\bf 10}}$.
}

\newpage

\section{Introduction}
The CDF collaboration has recently reported an updated result of the $W$ boson mass~\cite{CDF:2022hxs},
\begin{eqnarray}
(M_W)_{\rm CDF} = (80.4335 \pm 0.0094)\,{\rm GeV}.
\end{eqnarray}
When combined with previous results from LEP\,2, Tevatron, LHC and LHCb experiments, the average value obtained is as~\cite{deBlas:2022hdk}
\begin{eqnarray}
(M_W)_{\rm exp} =(80.4133 \pm 0.0080)\, {\rm GeV},
\end{eqnarray}
which deviates from the Standard Model (SM)  prediction~\cite{deBlas:2022hdk} by 6.5 $\sigma$:
\begin{eqnarray}
(M_W)_{\rm SM} = (80.3500 \pm 0.0056)\, {\rm GeV}.
\end{eqnarray}

If the CDF result is indeed confirmed in future experiments, new physics is required to explain this shift in the $W$ boson mass. With this in mind, many new physics scenarios for explaining this shift have been proposed (see, e.g., ~\cite{Dcruz:2022dao} and references therein). As pointed out in Ref.~\cite{Strumia:2022qkt} (see also \cite{Lu:2022bgw} for the global electroweak fit with the CDF result), if this anomaly is due to loop effects of new particles, it generically requires new particles lighter than a few hundred GeV if the size of the relevant coupling is similar to the $SU(2)_L$ gauge coupling. Futhermore, these light particles tend to be heavily constrained by the LHC and other existing experiments. Thus, a search for a tree-level explanation of the $W$ boson mass shift is warranted. Among the possible scenarios, the addition of an $SU(2)_L$ triplet Higgs field with zero hypercharge~\cite{DiLuzio:2022xns, Athron:2022isz, Perez:2022uil} is likely the most economical solution. In this scenario, the $SU(2)_L$ triplet Higgs field obtains a small vacuum expectation value (VEV) which contributes only to the $W$ boson mass~\cite{Ross:1975fq, Gunion:1989ci, Blank:1997qa, Forshaw:2003kh, Chen:2006pb, Chankowski:2006hs, Chivukula:2007koj}.\footnote{
See Refs.~\cite{Kribs:2007ac, Diessner:2014ksa, Diessner:2019ebm} for supersymmetric models with the triplet Higgs.
} 

Although this scenario is phenomenlogically viable, why is a triplet Higgs field this light necessary at the weak scale since it seems to have little effect on the weak scale other than to shift the $W$ boson mass. As a possible answers to this question, we propose a simple  Grand Unified theory (GUT). We will show that, if the triplet Higgs field is complex and has a mass of order a TeV scale, the three fundamental gauge coupling constants in SM unify rather precisely at around $10^{14}$\,GeV. A unification scale this low may seem problematic due to proton decay constraint. However, we will show that rapid proton decay can be avoided by introducing vector-like fermions which mix with the SM fermions through a GUT breaking VEV dependent mass terms. Importantly, the minimal model only requires one pair of ${\bf 10}$ and $\overline{{\bf 10}}$ fermions which mix with the $SU(2)_L$ doublet quark, $SU(2)_L$ singlet quark and lepton of the SM ${\bf 10}$.

\section{Weak Scale Complex $SU(2)_L$ Triplet Higgs Boson}
Here, we discuss the $W$ boson mass shift due to the addition of a complex zero hyperchage triplet Higgs boson of $SU(2)_L$. This model bears some resemblance to those discussed in~\cite{Athron:2022isz, Perez:2022uil}. From the perspective of the $W$ boson mass, the complex triplet adds very little.  However, phenomenologically it is different and as we will see later, this complex triplet has significant implications for grand unification. 

\subsection{$W$ Boson Mass }
In this section, we examine electroweak symmetry breaking with the addition of a complex triplet.  The Lagrangian we consider is 
\begin{eqnarray}
\mathcal{L} = 2 {\rm Tr}\left[ (D_\mu \Sigma_3)^\dag (D^\mu \Sigma_3) \right] - V(H, \Sigma_3), \label{eq:triplet}
\end{eqnarray}
where 
\begin{eqnarray}
D_\mu \Sigma_3 = \partial_\mu \Sigma_3 - i g_2 [W_\mu, \Sigma_3], \nonumber \\
\Sigma_3 = \frac{\tau^a}{2} \Sigma^a, \ \ W_\mu = \frac{\tau^a}{2} W_\mu^a.
\end{eqnarray}
Here, $\tau^a, a=1,2,3$ are Pauli matrices. The scalar potential is given by 
\begin{eqnarray}
 V(H, \Sigma_3) &=& - \mu_H^2 |H|^2  + \lambda_H |H|^4 +  A_{3 H} H^\dag \Sigma_3 H + h.c. \nonumber \\
&+&  2 \mu_3^2 {\rm Tr}( \Sigma_3^\dag \Sigma_3), \label{eq:Higgs_poten}
\end{eqnarray}
where $A_{3 H}$ is taken to be real and positive without a loss of generality, and $\lambda_H$, $\mu_H^2$ and $\mu_3^2$ are assumed to be positive. We have dropped the quartic terms other than $|H|^4$ for simplicity. The dropped terms do not significantly affect our conclusions. At the minimum of the potential, the Higgs doublet and Higgs triplet get the following vacuum expectation values: 
\begin{eqnarray}
\left< H \right> = (0, v)^T, \ \ 
\left<\Sigma_3\right> = 
\frac{1}{2}
\left(
\begin{array}{ccc}
v_T  & 0    \\
0  &  -v_T
\end{array}
\right).
\end{eqnarray}
From the minimization conditions, we get the conditions
\begin{eqnarray}
v^2 =(\mu_H^2 + A_{3 H} v_T)/(2\lambda_H), \ \ 
v_T =  \frac{A_{3 H} v^2}{2\mu_3^2}.
\end{eqnarray}
The $W$-boson mass receives an additional contribution at tree-level: 
\begin{eqnarray}
\delta M_W^2 = 2 g_2^2 v_T^2.
\end{eqnarray}
Therefore, to explain the $W$-boson mass anomaly at the $1 \sigma$ level, $v_T\approx 3.2$\,GeV is required.

The real parts of $\Sigma_3 \ni \sigma_3/\sqrt{2}$ and $H_0 \ni \sigma_0/\sqrt{2}$ are mixed due to the third term in Eq. (\ref{eq:Higgs_poten}). The mass matrix is given by
\begin{eqnarray}
V \ni \frac{1}{2}
\left(
\begin{array}{cc}
\sigma_0  &  \sigma_3 
\end{array}
\right)
\left(
\begin{array}{cc}
 4 \lambda_H v^2 & - A_{3 H} v   \\
- A_{3 H} v  &  \mu_3^2    
\end{array}
\right)
\left(
\begin{array}{c}
\sigma_0  \\
\sigma_3 
\end{array}
\right).
\end{eqnarray}
The eigenvalues of the masses are 
\begin{eqnarray}
m_{h}^2 &=& \frac{1}{2}\left[ \mu_3^2 + 4 \lambda_H v^2 - \sqrt{(\mu_3^2 - 4 \lambda_H v^2)^2 + 4 A_{3 H}^2 v^2} \right], \nonumber \\
m_{\sigma_H}^2 &=& \frac{1}{2}\left[ \mu_3^2 + 4 \lambda_H v^2 + \sqrt{(\mu_3^2 - 4 \lambda_H v^2)^2 + 4 A_{3 H}^2 v^2} \right], \nonumber \\
\tan 2\theta_h &=& \frac{2 A_{3H} v }{\mu_3^2-4 \lambda_H v^2},
\end{eqnarray}
where $\theta_h$ is a mixing angle. For $\mu_3 \gg v, A_{3_H}$ we can expand the masses as
\begin{eqnarray}
m_{h}^2 &\simeq& 4 \lambda_H v^2 - \frac{A_{3H}^2 v^2 }{\mu_3^2} + \mathcal{O}(v^4)
\nonumber \\ 
m_{\sigma_H}^2 &\simeq& \mu_3^2 + \frac{A_{3H}^2 v^2 }{\mu_3^2} + \mathcal{O}(v^4),
\end{eqnarray}
Therefore, under the condition that $v_T = A_{3H} v^2/(2\mu_3^2) \sim 2$\,GeV, and $\mu_3 \gtrsim 1$\, TeV, the mass of SM Higgs boson and its couplings to other particles are nearly unaffected. The other fields, a CP-odd Higgs, $A$, and two charged Higgs, $H_1^\pm$ and $H_2^\pm$ have masses of $\mu_3$, $\mu_3$ and $\sqrt{\mu_3^2 + \frac{A_{3H}^2 v^2}{\mu_3^2}}$, respectively. LHC constraints on charged Higgs boson mass from $t b$ production is at most around 1\,TeV~\cite{ATLAS:2021upq}. In our case, due to the suppressed mixing between $\Sigma_3$ and $H$, the constraint is much weaker (see Appendix~\ref{ap:charged_higgs}).

\subsection{Gauge Coupling Unification}
Here, we consider the effect on gauge coupling unification of a complex $SU(2)_L$ triplet Higgs boson field with a hypercharge of zero. When we consider the renormalization group equations (RGEs) for the gauge coupling constants, this triplet Higgs boson contributes only to the running of the $SU(2)_L$ gauge coupling. By varying the mass of this triplet Higgs boson, the renormalization group running of the weak gauge coupling, $g_2$, can be deflected making it possible that all three gauge couplings could be made to unify. 

\begin{figure}[htp]
\centering
\includegraphics[scale=0.52]{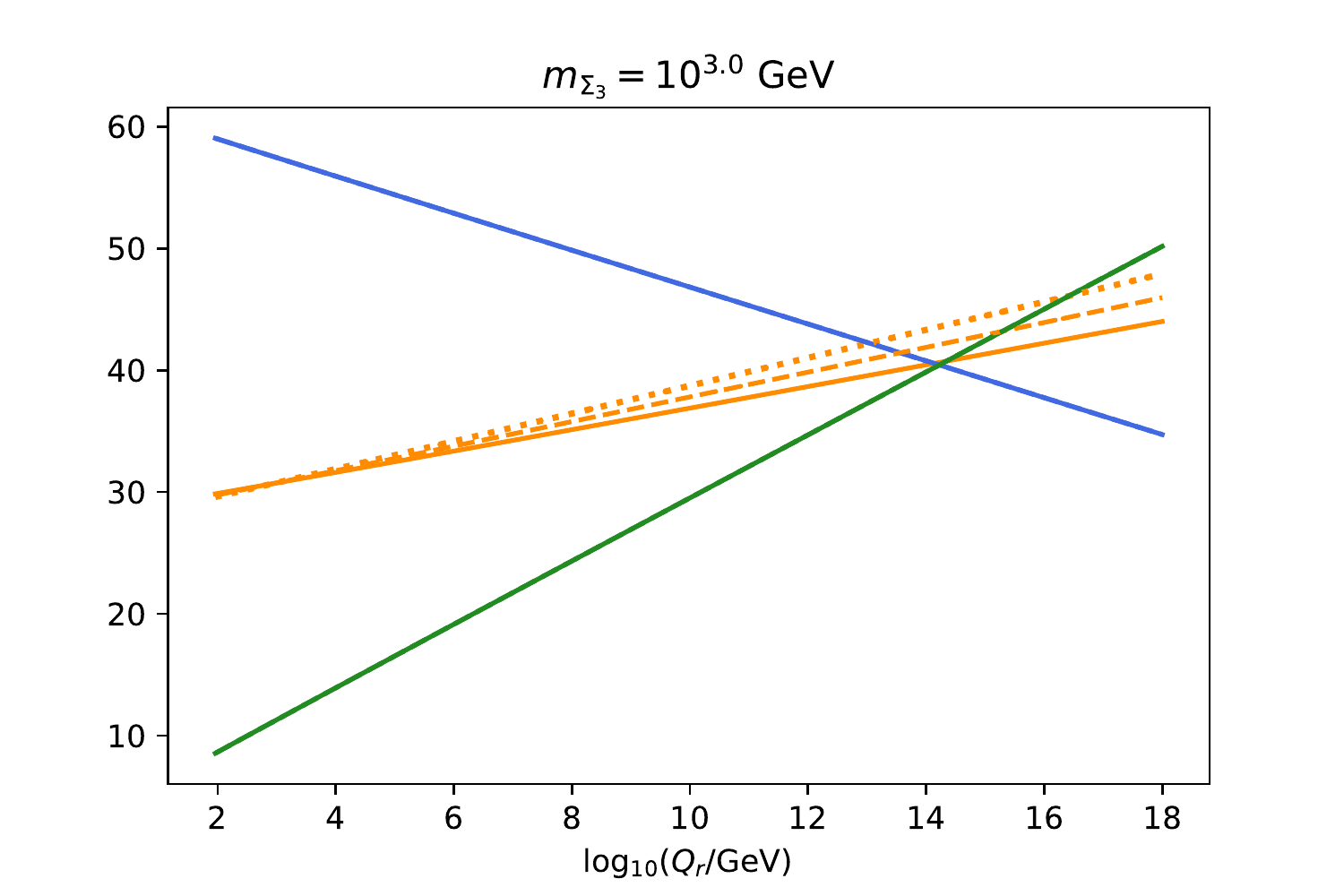}
\includegraphics[scale=0.52]{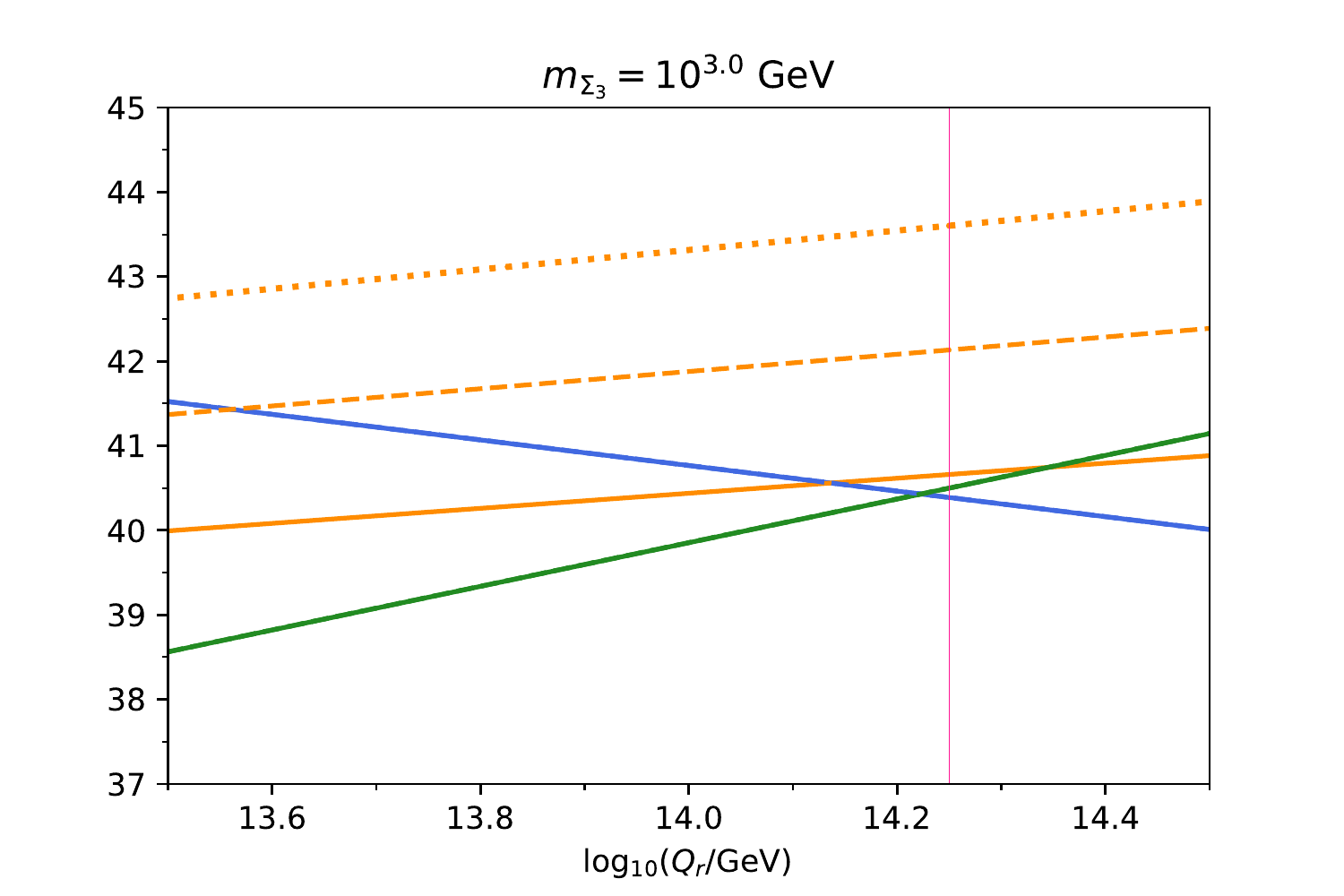}
\caption{Renormalzation group evolution of gauge coupling constants in SM (dotted), SM + one real triplet Higgs (dashed) and SM + one complex triplet Higgs (solid). The mass of the triplet Higgs is taken as 1\,TeV} 
\label{fig:1}
\end{figure}

In Fig.~\ref{fig:1}, we plot the running of $\alpha_{1,2,3}^{-1}$. The blue, orange and green lines show $\alpha_1^{-1}$, $\alpha_2^{-1}$ and $\alpha_3^{-1}$, respectively. The normalization of $U(1)_Y$ coupling is taken such that $g_1^2 = \frac{5}{3} g_Y^2$. To compute the renormalization group (RG) evolution, we use two-loop RGEs obtained using {\tt PyR@TE 3}~\cite{Sartore:2020gou}. The dotted, dashed and solid lines show the RG running for: the SM, the SM + one real triplet Higgs , and the SM + one complex triplet Higgs (i.e. two real triplets). As is clear from this figure, the unification becomes better for each triplet Higgs which is added. For the case with a complex triplet, two real triplets, the coupling unification is rather precise. In this case, the three gauge couplings meet at around $10^{14}$\,GeV. A prori, it seems natural to identified this scale as the GUT scale \footnote{Note that in the real triplet case, the differences among $\alpha_{i}^{-1}$ is quite large, and it is very difficult to be consistent with a $SU(5)$ GUT model even if we take into account the GUT scale threshold corrections. (See Appendix \ref{ap:real24}.) }. As we will discuss in more detail next, a GUT scale this low can be problematic from the perspective of proton decay.  

\subsection{Proton decay}
A serious drawback to our current model is the predicted proton lifetime. The prediction of the proton lifetime for minimal SU(5) with an $X$ gauge boson mass of order $10^{14}$ GeV is much shorter than the experimental lower bound. Thus, we need a mechanism to suppress the proton decay rate.  A naive estimate for the proton decay due to the $X$ gauge bosons is~\cite{Hisano:2012wq}   
\begin{eqnarray}
\tau_p=\frac{1}{\Gamma_p} \sim \frac{M_X^4}{\alpha_{\rm GUT}^2 m_p^5}=2.4\times 10^{34} {\rm yrs} \left(\frac{\frac{1}{40}}{\alpha_{\rm GUT}}\right)^2\left(\frac{M_X}{5.3\times 10^{15} {\rm GeV}}\right)^4
\end{eqnarray}
where the current limit on the proton lifetime is $2.4\times 10^{34}$ years~\cite{Super-Kamiokande:2020wjk}, $\alpha_{\rm GUT}$ is gauge couplings constant at the GUT scale, $m_p$ is the proton mass, and $M_X$ is the mass of the heavy SU(5) gauge bosons. As is clear from examining Fig. (\ref{fig:1}), the predicted lifetime of the above model is orders of magnitude too short.  

In this subsection, we show how the proton decay width can be suppressed by introducing vector-like fermions transforming as ${\bf 5}+\bar{\bf 5}$ and ${\bf 10}+\overline{\bf 10}$.\footnote{
See also Refs.~\cite{Ibe:2019ifm,Ibe:2022ock} for suppression of the proton decay amplitudes by the mixing with hidden vector-like leptons, which are charged under $U(2)_H$.
} The minimal cases requires a single pair ${\bf 10}+\overline{\bf 10}$. By mixing these vector like quarks with the usual quarks and leptons which are coupled to the Higgs fields, proton decay mediated by the heave gauge bosons can be suppressed. In order to realize the needed mixing, these vector like fermions need to get a contribution to their mass from the GUT breaking field.  This then allows us to realize a theory where the low-energy SM leptons and quarks come from different $SU(5)$ multiplets. This then gives a mixing angle suppression to all the interactions of the SM leptons and quarks. 

Now, we give the details of how to suppress proton decay using mixing. To minimal SU(5), we add a vector like pair that transforms as a ${\bf 10}+\overline{\bf 10}$ which we label as $\psi'_{10,i}$ and $\psi_{\overline{10},i}$, prime here indicate it is not the mass eigenstate. For now, we will only consider this model, which is sufficient to suppress proton decay. Later, we will mention other models involving the addition of a ${\bf 5}+{\bar {\bf 5}}$. 
\begin{eqnarray}
-\mathcal{L} &\ni& 
   \psi_{\overline{10} ,i} (  M_{10,i} + \lambda_{10,i} \left<\Sigma_{24 H}\right> ) {\bf 10}'_i +  \psi_{\overline{10} ,i}  ( M'_{10, i} +\lambda'_{10, i} \left<\Sigma_{24 H}\right> ) \psi'_{10,i}  + h.c., \label{eq:vec_mixing}
\end{eqnarray}
where would-be SM matter fields can be written in terms of $SU(3)_c \times SU(2)_L \times U(1)_Y$ components as
${\bf 10}'_i = (Q'_i,\bar{U}'_i,\bar{E}'_i)$,
and $\psi'_{10,i} = (\psi'_{Q_i}, \psi'_{\bar{U}_i},\psi'_{\bar{E}_i})$.
Also, we ignore flavor mixing just for simplicity. Taking $\left<\Sigma_{24 H}\right>=v_{\rm GUT}\, {\rm diag}(2,2,2,-3,-3)$, the above Lagrangian simplifies to 
\begin{eqnarray}
-\mathcal{L} &\ni&  M_{Q,i}\psi_{\bar Q_i}Q'_i+M'_{Q,i}\psi_{\bar Q_i}\psi'_{Q_i}\\
&& +M_{U,i}\psi_{U_i} {\bar U}'_i +M'_{U,i} \psi_{U_i}\psi'_{\bar{U}_i} +M_{E,i}\psi_{E_i}{\bar E}'_i +M'_{E,i} \psi_{E_i}\psi'_{\bar{E}_i} + h.c., \nonumber 
\end{eqnarray}
where 
\begin{eqnarray}
M_{Q,i} &=& M_{10} - \frac{1}{2} \lambda_{10,i} v_{\rm GUT} ,  \ \
M'_{Q,i} = M'_{10} - \frac{1}{2} \lambda'_{10,i} v_{\rm GUT} ,  \nonumber \\
M_{U,i} &=& M_{10} +2 \lambda_{10,i} v_{\rm GUT} ,   
\ \
M'_{U,i} = M'_{10} +2 \lambda'_{10,i} v_{\rm GUT} , 
\nonumber \\ 
M_{E,i} &=& M_{10} -3 \lambda_{10,i} v_{\rm GUT} ,
\ \  
M'_{E,i} = M'_{10} -3 \lambda'_{10,i} v_{\rm GUT},
\end{eqnarray}

Note that, although Eq.(\ref{eq:vec_mixing}) contains 3 pairs of vector-like fermions, to avoid the proton decay constraint, only one pair of $\psi_{\overline{10} ,1}$ and  $\psi'_{10,1}$ is minimally required. 

Due to the rank of the mass matrix for each of the $Q$-like, $U$-like, and $E$-like fields, only one combination of these fields, for each $i$, has a Dirac mass while the other combination is massless. 

The mixing matrices for these fermions then takes the form,
\begin{eqnarray}
\left(
\begin{array}{cc}
 \cos\theta_{(Q, U, E)_i} & \sin\theta_{(Q, U, E)_i}  \\
-\sin\theta_{(Q, U, E)_i} &   \cos\theta_{(Q, U, E)_i}  
\end{array}
\right)
\left(
\begin{array}{c}
(Q', \bar{U}', \bar{E}')_i  \\
 \psi'_{(Q, \bar{U}, \bar{E})_i} 
\end{array}
\right) = 
\left(
\begin{array}{c}
  \psi_{(Q, \bar{U}, \bar{E})_i}  \\
(Q, \bar{U}, \bar{E})_i
 \end{array}
\right),
\end{eqnarray}
where 
\begin{eqnarray}
 \cos\theta_{(Q, U, E)_i} = \frac{M_{(Q, U, E),i}}{\sqrt{ M^2_{(Q, U, E),i} + M^{'2}_{(Q, U, E),i}}}, \ \  
\sin\theta_{(Q, U, E)_i}   = \frac{M'_{(Q, U, E),i}}{\sqrt{ M^2_{(Q, U, E),i} + M^{'2}_{(Q, U, E),i}}  }. 
\end{eqnarray}
and $Q,\bar U,\bar E$ are the light states.

Importantly, we can realise $|M_{U,i}| \gg |M'_{U,i}|$ and $|M'_{Q,i}| \gg |M_{Q,i}|$ (or $|M_{U,i}| \ll |M'_{U,i}|$ and $|M'_{Q,i}| \ll |M_{Q,i}|$). In this case. the dimension 6 proton decay operators always contain ${\bar U}^\dag_1$ and $Q_1$. This allows us to suppress the proton decay rate by taking $\sin {\theta_{U_1}}$ and $\cos {\theta_{Q_1}}$ (or $\cos {\theta_{U_1}}$ and $\sin {\theta_{Q_1}}$) to be smaller than $10^{-4}$. For these mixing angles, $\bar{U}_1$ predominantly made of $\psi'_{\bar U,1}$, a component of $\psi'_{10,1}$, while $Q_1$ is predominantly composed of $Q'_1$, a component of $10'$, (or vice versa). Therefore the dimension 6 proton decay operators have coefficients smaller than $10^{-4}$, and we can suppress the proton decay rate. 

A very similar thing can be done to the $L$ and $\bar D$ by adding a pair of $5+\bar 5$ for each generation. In the GUT representation, the Lagrangian for these fields takes the form
\begin{eqnarray}
-\mathcal{L} &\ni& 
 {\bar {\bf 5}}'_i   (M_{5, i} + \lambda_{5, i} \left<\Sigma_{24 H}\right>) \psi_{5,i} +  \psi'_{\bar 5,i} (M'_{5, i} + \lambda'_{5, i} \left<\Sigma_{24 H}\right>) \psi_{5,i}  + h.c. \nonumber \\
\label{eq:vec_mixing_5}
\end{eqnarray}
where the would-be SM matter fields this time can be written in terms of 
$\bar{\bf 5}'_i = (\bar{D}'_i, L'_i)$ and
$\psi'_{\bar{5},i} = (\psi'_{\bar{D}_i}, \psi'_{L_i})$.

The mixing angles for these fields takes the form
\begin{eqnarray}
&&\left(
\begin{array}{cc}
 \cos\theta_{(L,D)_i} & \sin\theta_{(L,D)_i}  \\
-\sin\theta_{(L,D)_i}  &  \cos\theta_{(L,D)_i}  
\end{array}
\right)
\left(
\begin{array}{c}
  (L',\bar{D}')_i  \\
 \psi'_{(L,\bar{D})_i}  
\end{array}
\right) = 
\left(
\begin{array}{c}
 {\psi}_{(L, \bar{D})_i}  \\
  (L,\bar{D})_i  \\
\end{array}
\right),
\end{eqnarray}
where $\psi_{\bar{D}_i}$ and $ {\psi}_{L_i}$ are the heavy states, and $\bar{D}_i$ and $L_i$ are massless states. The mixing angles can be written as
\begin{eqnarray}
&&\cos\theta_{(L,D)_i} = \frac{M_{(L,D),i}}{\sqrt{M^2_{(L,D),i} + (M'_{(L,D),i})^2}}, \ \ 
\sin\theta_{(L,D),i} = \frac{M'_{(L,D),i}}{
\sqrt{M^2_{(L,D),i} + (M'_{(L,D),i})^2}}, \nonumber \\
\end{eqnarray}
where 
\begin{eqnarray}
M_{L,i} &=&  M_{5,i} -3 \lambda_{5, i} v_{\rm GUT}, 
\ \ 
M'_{L,i} = M'_{5,i} -3 \lambda'_{5, i} v_{\rm GUT}, \nonumber \\
M_{D,i} &=& M_{5,i} + 2 \lambda_{5, i} v_{\rm GUT}, 
\ \  M'_{D,i} = M'_{5,i} +  2 \lambda'_{5, i} v_{\rm GUT},
\end{eqnarray}
and the heavy states have masses,
\begin{eqnarray}
M_{({\tilde L},{\tilde D}),i} =  \sqrt{M_{(L,D),i}^2 + (M'_{(L,D),i})^2}.
\end{eqnarray}

This set up offers an alternative way to suppresses proton decay. However, small mixing angles for $L$ and $\bar D$ alone is insufficient to suppress proton decay and must be supplemented with mixing angle suppression of $\bar E$ and $Q$ as will be made clear below.

Using the above mixing angles, we estimate the life-time of the proton, in the channel $\tau(p \to e^+ \pi^0)$, to be~\cite{Hisano:2012wq}
\begin{eqnarray}
\tau(p \to e^+ \pi^0) \approx 3.3 \times 10^{27} {\rm yrs} \, A_{\rm mix}^{-1}(i=1)\, \left(\frac{M_X}{10^{14}{\rm GeV}}\right)^4 \left(\frac{g_5}{0.55}\right)^{-4},
\end{eqnarray}
where 
\begin{eqnarray}
A_{\rm mix}(i) &\simeq& (\cos\theta_{Q_1} \cos\theta_{U_1} + \sin\theta_{Q_1} \sin\theta_{U_1})^2 \nonumber \\
&\times&
\left[
(\cos\theta_{D_i} \cos\theta_{L_i} + \sin\theta_{D_i} \sin\theta_{L_i})^2  + (\cos\theta_{Q_i} \cos\theta_{E_i} + \sin\theta_{Q_i} \sin\theta_{E_i})^2
\right]. \label{eq:plife}
\end{eqnarray}
Another important decay mode, $\tau(p \to \mu^+ \pi^0)$, can be estimated from the above expression by taking $A_{\rm mix}(i=2)$. To avoid the experimental bound, $\tau(p \to e^+ \pi^0) > 2.4 \times 10^{34}\,{\rm yrs}$ and $\tau(p \to \mu^+ \pi^0) > 1.6 \times 10^{34}\,{\rm yrs}$~\cite{Super-Kamiokande:2020wjk},
we need to satisfy either of the following conditions,
\begin{eqnarray}
&&(a)\   |\sin\theta_{Q_1}|,|\cos\theta_{U_1}| \leq 10^{-4}{\rm \  or\ }  |\cos\theta_{Q_1}|,|\sin\theta_{U_1}| \leq 10^{-4}. \nonumber \\
&&(b)\  \left( |\sin\theta_{L_i}|,|\cos\theta_{D_i}| \leq 10^{-4}{\rm \  or\ } |\cos\theta_{L_i}|,|\sin\theta_{D_i}| \leq 10^{-4} \right) \nonumber \\
&& {\rm\  and \ } \left(  |\sin\theta_{E_i}|,|\cos\theta_{Q_i}| \leq 10^{-4} {\rm \ or \ } |\cos\theta_{E_i}|,|\sin\theta_{Q_i}| \leq 10^{-4} \right).
\end{eqnarray}
The condition $(a)$ is easier to be satisfied: in this case, we only need one pair of vector-like matter multiplet, $\psi'_{10,1}$ and $\psi_{\overline{10},1}$.

Because the way in which we suppress proton decay relies on GUT breaking mixing, its affect on the colored Higgs couplings turns out to be non-trivial. As we will now show, this leads to a large enhancement to proton decay which is mediated by the color Higgs boson. As a minimal example, let us consider the case where  $\sin\theta_{U_1}=\cos\theta_{Q_1}=10^{-4}$, and estimate the colored Higgs boson's contribution to proton decay. 
The Yukawa couplings are assumed to take the following  form\footnote{We take the other Yukawa couplings involving $\psi'_{10,i}$ and $\psi'_{\bar{5},i}$ to be zero for simplicity. However, the following conclusions still hold for appropriately chosen Yukawa couplings and mixings.}:
\begin{eqnarray}
-\mathcal{L} &\ni& \frac{1}{4} Y_{10,i} \delta_{ij}
{\bf 10}'_i {\bf 10}'_j H_5 + \sqrt{2} Y_{5,ij} {\bf \bar 5}_i {\bf 10}'_j  H_5^* + h.c.,
\end{eqnarray}
where ${\bf 10}'_i=({\bf 10}'_1, {\bf 10}_2, {\bf 10}_3)$, and we omit $SU(5)$ gauge indices.\footnote{
They can be written explicitly as
\begin{eqnarray}
{\bf 10}'_i {\bf 10}'_j H_5 \equiv \epsilon_{abcde}
({\bf 10}'_i)^{ab} ({\bf 10}'_j)^{cd} (H_5)^e, 
\ \ 
{\bf \bar 5}'_i {\bf 10}'_j  H_5^* \equiv ({\bf \bar 5}'_i)_a ({\bf 10}'_j)^{ab} (H_5^*)_b. \nonumber \\
\end{eqnarray}
}
Before we finish our discussion on colored Higgs bosons meadieated proton decay, we need to address how to realize unification of the Yukawa couplings. As is well known, the observed fermion masses are correctly explained by including higher dimensional operators~\cite{Ellis:1979fg,Panagiotakopoulos:1984wf,Bajc:2002pg},
\begin{eqnarray}
-\mathcal{L}_{{\rm dim}\,5} \ni
\sqrt{2} \frac{c_{ij}}{M_*} {\bf \bar 5}_i \Sigma_{24H} {\bf 10}'_j  H_5^* + h.c.,
\end{eqnarray}
where $M_*$ is the cut-off scale. 
The SM Yukawa interactions are then 
\begin{eqnarray}
-\mathcal{L}_{SM} \ni (Y_{u})_i Q_i \cdot H \bar U_i+(Y_{d})_{ij} {\bar D}_i H^\dagger Q_j 
+ (Y_{e})_{ij} H^\dagger L_i \bar E_j + h.c.
\end{eqnarray}
where
\begin{eqnarray}
Y_{u} &\approx&Y_{10,1}\sin\theta_{U_1}, \ \ 
Y_{c} = Y_{10,2}, \ \  Y_{t} = Y_{10,3}, 
 \nonumber \\
(Y_{d})_{ij} &\approx& Y_{5,ij} + 2 c_{ij} x_{\rm GUT}
= ({\hat Y}_d V_{\rm CKM}^\dag)_{ij} 
 \nonumber \\
(Y_{e})_{ij} &=& [({\hat Y}_d V_{\rm CKM}^\dag)_{ij}-5 c_{ij} x_{\rm GUT})] (1 + (\sin\theta_{E_1}-1) \delta_{1 j}) \nonumber \\
&=& (V_L^* {\hat Y}_{e} U_e^T)_{ij},
\end{eqnarray}
and
\begin{eqnarray}
x_{\rm GUT} = \frac{v_{\rm GUT}}{M_*}, \ \ 
{\hat Y}_d = {\rm diag}(m_d, m_s, m_b)/v, \ \ 
{\hat Y}_e = {\rm diag}(m_e, m_\mu, m_\tau)/v.
\end{eqnarray}
In the minimal setup, the cut-off scale needs to be $M_* \sim 10^{16}$\,GeV to explain $m_{\tau}/m_{b} \approx 1.5$ (see, e.g., Ref.~\cite{Bora:2012tx}). However, the cut-off scale can easily be increased to the (reduced) Planck mass scale with a slight extension as we will show at the end of this subsection.

The interactions involving the colored Higgs can be written by
\begin{eqnarray}
-\mathcal{L} &\ni& Y_{E_iU_i} \delta_{ij} H_C\bar E_i \bar U_j+\frac{1}{2} Y_{Q_iQ_i} \delta_{ij} H_C  \left(Q_i\cdot Q_j\right) \\ \nonumber  &+&Y_{D_i U_j}\bar D_i \bar U_j H_C^* +Y_{L_i Q_j}\left(L_i\cdot Q_j\right) H_C^* +{\rm h.c.} 
\end{eqnarray}
where 
\begin{eqnarray}
Y_{E_i U_i} &\approx&  (Y_u, Y_c, Y_t), \ \
Y_{Q_i Q_i} \approx (Y_u/\sin\theta_{U_1}, Y_c, Y_t), \nonumber \\
Y_{D_i U_j} &\approx& (Y_{d})_{ij}, \ \ 
Y_{L_i Q_j} \approx \frac{(Y_{e})_{ij}}{1 + (\sin\theta_{E_1}-1) \delta_{1 j}}. 
\end{eqnarray}
By taking $\sin\theta_{E_1} \approx 1$ for simplicity, the only difference in these coplings from the minimal $SU(5)$ case is $Y_{Q_1 Q_1}$, which is enhanced by $\sin\theta^{-1}_{U_1} =10^4$. Since this coupling is important is mediating proton decay via the colored Higgs boson, the bounds on the color mass, $m_{H_C}$, from proton decay are significantly stronger.

We now perform a somewhat rough estimate of the colored Higgs boson mass needed to evade proton decay constraints. For simplicity, we will assume that $V_L$ is the identity.  In this case, we can go to the mass eigenstate basis by $\bar{E} \to U_e^* \bar{E}$ and $d_L \to V_{\rm CKM} d_L\, (d_L \in Q)$, while $L$, $\bar D$, and $u_L$ are already in their mass eigenstate basis.  
By integrating out the colored Higgs, we generate the follow proton decay operator which gives one of the dominant contributions to the decay mode $p\to K^+\overline{\nu}$
\begin{eqnarray}
-\mathcal{L} &\ni& \left(\frac{Y_{Q_1Q_1}Y_{L_3Q_i}(V_{CKM})_{i2}}{m_{H_C}^2}\right)(ud)_Ls_L\nu_\tau ~.
\end{eqnarray}
There are other contribution to this decay and decays to other nuetrino flavors but they are at most similar in size and and the larger contributions do not add constructively. Thus, they do not have a significant effect.  We will assume $U_e^T \sim V_{CKM}^{\dagger}$. We also ignore the RG running of the Wilson coefficient, since we are just looking for a rough estimate, and apply the hadron matrix elements calculated on the lattice. For the operator of interest, the lattice matrix element can be found in~\cite{Aoki:2017puj} and is 
\begin{eqnarray}
\langle K^+\left| (ud)_Ls_L\right|p \rangle =  0.139 \,{\rm GeV^2}
\end{eqnarray}
The amplitude for this matrix element can then be written as 
\begin{eqnarray}
A(p\to K^+\overline{\nu}_{\tau}) &=& \langle K^+\left| (ud)_Ls_L\right|p \rangle \left(\frac{Y_{Q_1Q_1}Y_{L_3Q_i}(V_{CKM})_{i2}}{m_{H_C}^2}\right)\\ \nonumber
&\simeq& \langle K^+\left| (ud)_Ls_L\right|p \rangle \frac{y_\tau(M_{GUT}) y_u(M_{GUT}) (V_{CKM}^*)_{\rm ts}}{\sin\theta_{U_1}m_{H_C}^2}
\end{eqnarray}
where in the last equality we have approximated $(U_e^T V_{CKM})_{32}\sim (V_{CKM})_{ts}$ so as to capture a more generic set of models. For more generic quark mixing, this decay mode can have additional mixing angles associated with. The decay width from this amplitude is then 
\begin{eqnarray}
\Gamma_{p\to K^+\overline{\nu}_{\tau}}= \frac{m_p}{32\pi}\left(1-\frac{m_K^2}{m_p^2}\right)^2\left|A(p\to K^+\overline{\nu}_{\tau})\right|^2
\end{eqnarray}
where $m_K$ is the Kaon mass.  Using this expression, we find that the proton lifetime is
\begin{eqnarray}
\tau_p=\frac{1}{\Gamma_{p\to K^+\overline{\nu}_{\tau}}} = 9.4\times 10^{33}{\rm  yrs} \left(\frac{m_{H_C}}{2\times 10^{13}{\rm GeV}}\right)^4.
\end{eqnarray}
As can be seen from this expression, the life-time is longer than the experimental bound \cite{Super-Kamiokande_knu} for $m_{H_C}>\mathcal{O}(10^{13})\,{\rm GeV}$.

Now, there are other somewhat large contributions to proton decay, e.g. decays to $\nu_\mu$.  Their is also some enhancement of Wilson coefficient from RG running. However, all of these have no more than an order one effect and the lifetime scales as the colored Higgs to the fourth power. Thus, this estimate on the colored Higgs mass should be fairly robust. As we will discuss later, colored Higgs masses larger than this can easily be realized in our model and so this estimate is sufficient for our purposes.

Lastly, let us estimate the contributions to the gauge coupling constants from the additional heavy fermions. Their contribution to the gauge couplings at the one-loop level is 
\begin{eqnarray}
\Delta \alpha_1^{-1} (M_X) &=& 
-\sum_i \frac{2}{3}\frac{1}{2\pi} \left[
 \frac{3}{5}\ln\frac{M_X}{M_{{\tilde L}_i}} 
+ \frac{2}{5}\ln\frac{M_X}{M_{{\tilde D}_i}} 
+ \frac{1}{5} \ln\frac{M_X}{M_{{\tilde Q}_i}}
+ \frac{8}{5} \ln\frac{M_X}{M_{{\tilde U}_i}}
+ \frac{6}{5} \ln\frac{M_X}{M_{{\tilde E}_i}}
\right].
\nonumber \\
\Delta \alpha_2^{-1} (M_X) &=& -\sum_i \frac{2}{3}\frac{1}{2\pi} \left[
\ln\frac{M_X}{M_{{\tilde L}_i}} 
+ 3 \ln\frac{M_X}{M_{{\tilde Q}_i}}
\right]. \nonumber \\
\Delta \alpha_3^{-1} (M_X) &=& -\sum_i \frac{2}{3}\frac{1}{2\pi} \left[
\ln\frac{M_X}{M_{{\tilde D}_i}} 
+ 2 \ln\frac{M_X}{M_{{\tilde Q}_i}}
+ \ln\frac{M_X}{M_{{\tilde U}_i}}
\right].
\end{eqnarray}
First, lets consider the minimal model which avoids proton decay constraints. This model requires one additional pair of vector-like fermions, which mixes with first generation field ${\bf 10}'_1 = (Q'_1, \bar{U}'_1, \bar{E}'_1)$. Their contribution to the gauge couplings from splitting their masses can be suppressed by taking 
\begin{eqnarray}
M'_{U,1} \simeq 0, \ \ 
M_{Q,i} \simeq 0, \ \ 
|M_{\tilde{U}_1}| = |M'_{\tilde{Q},1}| = \frac{5}{2} |\lambda_{10,1}| v_{\rm GUT}, \ \  M_{\tilde{E}_1}=\frac{5\sqrt{5}}{2} |\lambda_{10,1}| v_{\rm GUT}.
\end{eqnarray}
Since any universal shift of all three gauge couplings has no effect on unification, we need only consider the non-universal shift. For the mass hierarchies above, we get a shift of  
\begin{eqnarray}
\Delta \alpha_1^{-1}-\Delta \alpha_2^{-1}=\Delta \alpha_1^{-1}-\Delta \alpha_3^{-1} = \frac{1}{5\pi}\ln(5) \approx 0.10,
\end{eqnarray}
which can be easily be accounted for by the threshold corrections which we discuss in the next subsection.\footnote{
In fact, this shift gives slightly better unification (see Fig.~\ref{fig:1}).
}

If we add an additional pair of vector-like fermions, we can explain the discrepancy in the the SM Yukawa couplings for a cut-off scale of the Planck scale. This can be done by the introduction of a $\psi_{5,3}$ and $\psi'_{{\bar 5},3}$, which mix with the $\bar{\bf 5}'_3$. Then, we get $(Y_e)_{33}/(Y_d)_{33} \simeq (Y_e)_{32}/(Y_d)_{32} \simeq \sin\theta_{L_3}/\sin\theta_{D_3}$. In this case, we can still take the parameters such that the loop-level contribution of the heavy states to the SM gauge couplings is universal. This can be accomplished by taking $M_{\tilde{D}_3}=M_{\tilde{L}_3}$ for the physical masses. This still afford the freedom to choose $\sin_{\theta_{D_3}}$ and $\sin_{\theta_{L_3}}$. Note that as long as the vector-like fermions are a pair of ${\bf 5} + \bar{\bf 5}$, we can take equal physical masses and arbitrary mixing angles.
Thus, for the sake of simplicity, we do not consider the corrections to the gauge couplings from the vector-like fermions.

\subsection{$SU(5)$ grand unification}

We now show that a complex $SU(2)_L$ triplet Higgs boson with a mass of around a TeV is consistent with grand unification. Here, we consider a minimal extension of minimal $SU(5)$ where $SU(5)$ is broken to the SM gauge group by the VEV of a real scalar transforming as a ${\bf 24}$ under the $SU(5)$ gauge group. In addition to this  ${\bf 24}$ Higgs, we introduce a complex scalar transforming as ${\bf 24}$. This complex scalar boson ${\bf 24}$ contains the complex $SU(2)_L$ triplet Higgs whose VEV can explain the observed mass shift the $W$ boson.\footnote{If we assume a product group unification model based on $SU(5) \times SU(2)_H \times U(1)_H$~\cite{Yanagida:1994vq, Hotta:1995cd, Hotta:1996pn, Ibe:2003ys}, the presence of the light $SU(2)_L$ triplet boson can be easily explained: the triplet Higgs boson can be originally regarded as a $SU(2)_H$ triplet field, and after the symmetry breaking, $SU(5) \supset SU(2) \times SU(2)_H \to SU(2)_L$, this field becomes the $SU(2)_L$ triplet.}  Details of the Higgs potential are presented in Appendix~\ref{ap:model}.

Unification of the gauge coupling depends on loop effects from all particles in incomplete representations below the GUT scale. In this model, the releveant particles are $X$ gauge boson, $\Sigma_{8 H}, \Sigma_{3 H}, H_C, \Sigma_8$ and $X_{1,2}({\bf 3},{\bf 2},-5/6)$. The first four fields are constituents of the minimal $SU(5)$; $\Sigma_{8 H}$ and $\Sigma_{3 H}$ are real scalars having charges of $({\bf 8},{\bf 1},0)$ and $({\bf 1},{\bf 3},0)$ for the SM gauge symmetries $SU(3)_c \times SU(2)_L \times U(1)_Y$. They are included in the GUT breaking Higgs and $H_C$ is the colored Higgs. The adjoint field $\Sigma_8$ $({\bf 8},{\bf 1},0)$ and $X_{1,2}({\bf 3},{\bf 2},-5/6)$ come from the complex scalar transforms as ${\bf 24}$ under $SU(5)$, which contains the triplet Higgs responsible for explaining the $W$ boson mass anomaly. 

By matching the gauge coupling constants of the full theory (minimal $SU(5)$ + ${\bf 24}$ scalar) and the effective theory (SM + triplet Higgs) at the $X$ gauge boson mass scale, the gauge couplings are written as
\begin{eqnarray}
\tilde{\alpha}_1^{-1} (M_X) &=&  \alpha_1^{-1}(M_X)- \frac{b_{1,H_C}}{2\pi} \ln \frac{M_X}{m_{H_C}}- \frac{b_{1,X_1}}{2\pi} \ln \frac{M_X}{m_{X_1}}-
\frac{b_{1,X_2}}{2\pi} \ln \frac{M_X}{m_{X_2}},
\nonumber \\
\tilde{\alpha}_2^{-1}(M_X) &=& \alpha_2^{-1}(M_X)
- \frac{b_{2,X_1}}{2\pi} \ln \frac{M_X}{m_{X_1}}-
\frac{b_{2,X_2}}{2\pi} \ln \frac{M_X}{m_{X_2}}
-\frac{b_{2,\Sigma_{3 H}}}{2\pi}\ln \frac{M_X}{2 m_{\Sigma_{8 H}}}
, \nonumber \\
\tilde{\alpha}_3^{-1}(M_X) &=& \alpha_3^{-1}(M_X)- \frac{b_{3,H_C}}{2\pi} \ln \frac{M_X}{m_{H_C}}- \frac{b_{3,X_1}}{2\pi} \ln \frac{M_X}{m_{X_1}}-
\frac{b_{3,X_2}}{2\pi} \ln \frac{M_X}{m_{X_2}}, \nonumber \\
&-& \frac{b_{3,\Sigma_{8 H}}}{2\pi}\ln \frac{M_X}{m_{\Sigma_{8 H}}}
- \frac{b_{3,\Sigma_{8}}}{2\pi}\ln \frac{M_X}{m_{\Sigma_{8}}},
\end{eqnarray}
where the coefficients of the one-loop beta-functions are given by
\begin{eqnarray}
&&b_{1,{H_C}}=1/15, \ \ b_{1,X_1}=5/6,  \ \ b_{1,X_2}=5/6,  \nonumber \\
&& b_{2,X_1}=1/2,  \ \ b_{2,X_2}=1/2, \ \ b_{2,\Sigma_{3H}}=1/3, \nonumber \\
&& b_{3,{H_C}}=1/6, \ \ b_{3,X_1}=1/3,  \ \ b_{3,X_2}=1/3, \ \  b_{3, \Sigma_8}=1, \ \ b_{3, \Sigma_{8 H}}=1/2. 
\end{eqnarray}
Here, ${\alpha}_i^{-1}(M_X)$ is a coupling constant evaluated from low-energy using the two-loop RGEs,
and $\tilde{\alpha}_{i}^{-1}(Q)$ evolves from $M_X$ to some higher energy scale according to a common beta-function, which is calculated in the full theory without the $SU(5)$ symmetry breaking. 
Therefore, we require that \begin{eqnarray}
\tilde{\alpha}_1^{-1}(M_X)= \tilde{\alpha}_2^{-1}(M_X) = \tilde{\alpha}_3^{-1}(M_X),
\end{eqnarray}
which determines the masses of GUT scale particles. We make the follow definitions for later simplicity: 
\begin{eqnarray}
k_{X_1} &=& \frac{1}{2\pi} \ln \frac{M_X}{m_{X_1}}, \ \ 
k_{X_2} = \frac{1}{2\pi} \ln \frac{M_X}{m_{X_2}}, \ \ 
k_{H_C} = \frac{1}{2\pi} \ln \frac{M_X}{m_{H_C}},
\nonumber \\
k_{8} &=& \frac{1}{2\pi} \ln \frac{M_X}{m_{\Sigma_8}}, \ \
k_{8H} = \frac{1}{2\pi} \ln \frac{M_X}{m_{\Sigma_{8H}}}~.
\end{eqnarray}
The conditions on the masses to obtain gauge coupling unification is then written as 
\begin{eqnarray}
k_{8H} &=& \frac{1}{2}\left[-5 {\alpha}_1^{-1}(M_X) + 3{\alpha}_2^{-1}(M_X)+2 {\alpha}_3^{-1}(M_X) \right]  - k_{8} + (k_{X_1} + k_{X_2}) + \frac{1}{4\pi} \ln 2, \nonumber \\
k_{H_C} &=& \frac{5}{4} \left[ 2 {\alpha}_1^{-1}(M_X) - 6  {\alpha}_2^{-1}(M_X)+4 {\alpha}_3^{-1}(M_X) - 4 k_8 
- \frac{1}{\pi} \ln 2 \right], \label{eq:gut_corr} 
\end{eqnarray}
where 
\begin{eqnarray}
k_{X_2} = -\frac{1}{4\pi}
\ln \left( e^{-4\pi k_8} + \frac{2}{3}e^{-4\pi k_{X_1}}
\right).
\end{eqnarray}
The pink vertical line in Fig.~\ref{fig:1} represents a possible $X$ gauge boson mass. On this line, the solution to Eq.(\ref{eq:gut_corr}), for instance, has $m_{H_C} \sim 10 M_X$ and $m_{X_1}, m_{X_2}, m_{\Sigma_{8}}, m_{\Sigma_{8H}},m_{\Sigma_{3H}}$ are all within the range $1/12 M_X$ and $12 M_X$.

\section{Conclusion}
We have shown that the $SU(2)_L$ triplet Higgs suggested by the CDF $W$-boson mass anomaly, significantly improve the gauge coupling unification compared to the SM case if the triplet Higgs is a complex field and exists around the TeV scale. This leads to the three SM gauge couplings unifying rather precisely at around $10^{14}$\,GeV. 
The light $SU(2)_L$ triplet Higgs required by the gauge coupling unification can be realized consistently within the framework of $SU(5)$ grand unified theory (see Appendix \ref{ap:model}). This complex triplet Higgs contains one CP-even Heavy Higgs, one CP-odd Higgs and two charged Higgs bosons, which could be the smoking gun single of this scenario.

Although the unification scale around $10^{14}$\,GeV is too low, in the usual sense, leading to significant proton decay constraints, we have shown that the constrains can be avoided by introducing additional vector-like fermions which mix with the SM fermions through an SU(5) breaking mass term. Importantly, the minimal requirement is quite simple and only requires the addition of a single pair of ${\bf 10}$ and $\overline{{\bf 10}}$ fermions to mix with the first generation ${\bf 10}$ matter multiplet. To get enough suppression in the proton decay rate, the $SU(2)_L$ singlet quark should have significant mixing with the vector-like fermion while $SU(2)$ doublet quark should have almost zero mixing with it (or vice versa). Interestingly, this leads to a suppression in the proton decay mediated by $X$ gauge bosons but leads to a significant enhancement in the proton decay through the colored Higgs boson. This means that if nature is realized by this minimal model, it is bound to show up in proton decay experiments eventually. 

Although this model has some additional fine tuning, the fine-tuning of the fermion masses is similar in nature to the doublet-triplet splitting present in all GUT models. Since the fine-tuning for all the fields in our model, including the the light complex $SU(2)_L$ triplet, are similar in design to the doublet-triplet splitting, it is possible that all the required tuning of this GUT theory is solved by a single mechanism, e.g. product group unification scenarios~\cite{Yanagida:1994vq, Hotta:1995cd, Hotta:1996pn, Ibe:2003ys}.

\section*{Acknowledgments}
N. Y. is supported by a start-up grant from Zhejiang University. J. L. E. is supported by a start-up grant from Shanghai Jiao-Tong University.
T. T. Y. is supported in part by the China Grant for Talent Scientific Start-Up Project and the JSPS Grant-in-Aid for Scientific Research No.~19H05810 and by World Premier International Research Center Initiative (WPI Initiative), MEXT, Japan.

\appendix

\section{Charged Higgs masses and mixing} \label{ap:charged_higgs}
To discuss the charged Higgs masses and mixing, let us expand $H$ and $\Sigma_3$ as
\begin{eqnarray}
H&=&(H^+, v)^T, \nonumber \\
\Sigma_3 &=& \frac{1}{2}
\left(
\begin{array}{cc}
 v_T & \sqrt{2} H_1^+  \\
\sqrt{2} H_2^-  &   -v_T
\end{array}
\right) \nonumber \\
\Sigma_3^\dag &=& \frac{1}{2}
\left(
\begin{array}{cc}
 v_T & \sqrt{2} H_2^+  \\
\sqrt{2} H_2^-  &   v_T
\end{array}
\right),
\end{eqnarray}
The mass terms for the charged Higgs bosons are written as
\begin{eqnarray}
V_{\rm charged} &=& \mu_3^2 (H_1^+ H_1^- + H_2^+ H_2^-) + \frac{A_{3H} v}{\sqrt{2}} (H_1^+ + H_2^+) H^-
+ \frac{A_{3H} v}{\sqrt{2}} H^+ (H_1^- + H_2^-) \nonumber \\
&+&\frac{1}{\mu_3^2} (A_{3H}^2 v^2) H^+ H^-
\end{eqnarray}
By taking
\begin{eqnarray}
H_a^{\pm} = \frac{1}{\sqrt{2}} (H_1^{\pm} + H_2^\pm), \ \ H_b^{\pm} = \frac{1}{\sqrt{2}} (-H_1^{\pm} + H_2^\pm),
\end{eqnarray}
the mass terms are then written as
\begin{eqnarray}
V_{\rm charged} &=& \mu_3^2 (H_a^+ H_a^- + H_b^+ H_b^-) + A_{3H} v H_a^+ H^-
+ A_{3H} v H^+ H_a^- \nonumber \\
&+&\frac{1}{\mu_3^2} (A_{3H}^2 v^2) H^+ H^-
\end{eqnarray}
and a linear combinations of $H_a^\pm$ and $H^\pm$ is the Goldstone boson, $G^\pm$.
\begin{eqnarray}
(G^+ \tilde {H}_a^+)^T = 
\left(
\begin{array}{ccc}
 \cos \theta_+ &  \sin\theta_+     \\
 -\sin \theta_+ &  \cos\theta_+    
\end{array}
\right) (H^+ H_a^+)^T,
\end{eqnarray}
and ${\tilde H}_a^+$ has a mass squared $\mu_3^2 + A_{3H}^2 v^2/\mu_3^2$. The mixing angle is again highly suppressed:
\begin{eqnarray}
\tan 2 \theta_+ = \frac{2 A_{3H} \mu_3^2 v}{-\mu_3^4+A_{3H}^2 v^2}.
\end{eqnarray}

\section{A GUT model for triplet Higgs} \label{ap:model}
Let us first consider the $SU(5)$ breaking potential for a real adjoint Higgs. The potential can be written by
\begin{eqnarray}
V \ni - \mu_{24H}^2 {\rm Tr} (\Sigma_{24 H}^2) 
+ \lambda_{1H} {\rm Tr} (\Sigma_{24 H}^2)^2
+ \lambda_{2H} {\rm Tr} (\Sigma_{24 H}^4),
\end{eqnarray}
where we have dropped a cubic term for simplicity.  
The minimum of the potential, $\left<\Sigma_{24_H}\right>=v_{\rm GUT}\, {\rm diag}(2,2,2,-3,-3)$, is obtained with
\begin{eqnarray}
v_{\rm GUT}^2 = \mu_{24}^2/(60 \lambda_{1H} + 14 \lambda_{2H}), \ \ 30 \lambda_{1H} + 7 \lambda_{2H} > 0.
\end{eqnarray}
The non-Goldstone modes $\Sigma_{3H}$ and $\Sigma_{8H}$ have masses
\begin{eqnarray}
m_{\Sigma_{3H}}^2 = 80 \lambda_{2H} v_{\rm GUT}^2, \ \ m_{\Sigma_{8H}}^2 = 20\lambda_{2H} v_{\rm GUT}^2,
\end{eqnarray}
Note that $m_{\Sigma_{3H}}^2/m_{\Sigma_{8H}}^2=4$, which does not change even if the cubic term is included. The singlet state in $\Sigma_{24 H}$ has a mass $2 \mu_{24H}$. 

Now we add the complex $24$, $\Sigma_{24}$, which contains the $SU(2)_L$ triplet Higgs shifting the $W$ boson mass:
\begin{eqnarray}
\Sigma_{24} = 
\left(
\begin{array}{ccc}
\Sigma_3  &   X_1/\sqrt{2}   \\
 X_2^\dag/\sqrt{2} &   \Sigma_8
\end{array}
\right)~.
\end{eqnarray}
Here, $X_1$ and $X_2$ transform $({\bf 3},{\bf 2},-5/6)$ under $SU(3)_c \times SU(2)_L \times U(1)_Y$. The $H^\dag \Sigma_3 H$ we consider in the main text originally comes from $H_5^\dag \Sigma_{24} H_5$. If we assume that the global $U(1)$ symmetry $\Sigma_{24} \to e^{i \alpha}\Sigma_{24}$ is only broken by $H_5^\dag \Sigma_{24} H_5$, the scalar potential giving the mass terms for $\Sigma_{24}$ can be written as
\begin{eqnarray}
V &\ni& 2\mu_{24}^2 {\rm Tr} (\Sigma_{24}^\dag \Sigma_{24}) 
+ 2A_1 {\rm Tr} (\Sigma_{24 H} \Sigma_{24}^\dag \Sigma_{24})
+ 2A_2 {\rm Tr} (\Sigma_{24}^\dag \Sigma_{24 H} \Sigma_{24}) \nonumber \\
&+& \lambda_{1} {\rm Tr} (\Sigma_{24 H}^2) {\rm Tr}(\Sigma_{24}^\dag \Sigma_{24})
+ 2\lambda_{2} {\rm Tr} (\Sigma_{24 H}^2 \Sigma_{24}^\dag \Sigma_{24}) 
+ 2\lambda_3 {\rm Tr} (\Sigma_{24 H} \Sigma_{24}^\dag \Sigma_{24 H} \Sigma_{24}).
\end{eqnarray}
By taking $\lambda_1=0$ (the contribution from non-zero $\lambda_1$ can be absorbed by the shift of $\mu_{24}^2$) and $A_2=0$ for simplicity, the mass parameters read
\begin{eqnarray}
m_{\Sigma_8}^2 &=& \mu_{24}^2 +2 A_1 v_{\rm GUT} + 4( \lambda_2 +  \lambda_3) v_{\rm GUT}^2, \nonumber \\
m_{\Sigma_3}^2 &=& \mu_{24}^2 -3 A_1 v_{\rm GUT} + 9(\lambda_2 + \lambda_3) v_{\rm GUT}^2 \nonumber \\
&=& m_{\Sigma_8}^2- 5 A_1 v_{\rm GUT}  + 5(\lambda_2 + \lambda_3) v_{\rm GUT}^2, \nonumber \\
m_{X_1}^2 &=& \mu_{24}^2 -3 A_1 v_{\rm GUT} + (9 \lambda_2 -6 \lambda_3) v_{\rm GUT}^2 \nonumber \\
&=& m_{\Sigma_3}^2 -15 \lambda_3 v_{\rm GUT}^2 \simeq -15 \lambda_3 v_{\rm GUT}^2, \nonumber \\
m_{X_2}^2 &=& \mu_{24}^2 +2 A_1 v_{\rm GUT} + (4 \lambda_2 -6 \lambda_3) v_{\rm GUT}^2 \nonumber \\
&=& m_{\Sigma_8}^2 -10 \lambda_3 v_{\rm GUT}^2 \simeq m_{\Sigma_8}^2 + \frac{2}{3} m_{X_1}^2.
\end{eqnarray}
By allowing the fine-tuning, $\lambda_3 \simeq -\lambda_2$ and $m_{\Sigma_8} \simeq 5 A_1 v_{\rm GUT}$, we can have $m_{\Sigma_3} \sim 1\,{\rm TeV}$ and, $m_{X_1} \sim m_{X_2} \sim m_{\Sigma_8} \sim M_X$. 

The singlet state $({\bf 1},{\bf 1},0)$ has a mass squared,  $\mu_{24}^2-A_1v_{\rm GUT} + 7(( \lambda_2 +  \lambda_3) v_{\rm GUT}^2) \simeq$ $2 A_1 v_{\rm GUT}$, which is also of the order of the GUT scale. It is, therefore, also irreverent for electroweak symmetry breaking and the $W$-boson mass anomaly.

\section{Real $24$ case} \label{ap:real24}
Here, we consider a model with only a real triplet coming from a real $24$. In this case, the GUT scale particles are 
$\Sigma_{3H},\Sigma_{8H},H_C$, $X({\bf 3},{\bf 2},-5/6)$ and $\Sigma_8$. By matching the gauge coupling constants at the $X$ gauge boson mass scale, the loop corrected gauge couplings can be written as
\begin{eqnarray}
\tilde{\alpha}_1^{-1} (M_X) &=&  \alpha_1^{-1}(M_X)- \frac{b_{1,H_C}}{2\pi} \ln \frac{M_X}{m_{H_C}}- \frac{b_{1,X}}{2\pi} \ln \frac{M_X}{m_{X}}-
\nonumber \\
\tilde{\alpha}_2^{-1}(M_X) &=& \alpha_2^{-1}(M_X)
- \frac{b_{2,X}}{2\pi} \ln \frac{M_X}{m_{X}}
-\frac{b_{2,\Sigma_{3 H}}}{2\pi}\ln \frac{M_X}{2 m_{\Sigma_{8 H}}}
, \nonumber \\
\tilde{\alpha}_3^{-1}(M_X) &=& \alpha_3^{-1}(M_X)- \frac{b_{3,H_C}}{2\pi} \ln \frac{M_X}{m_{H_C}}- \frac{b_{3,X}}{2\pi} \ln \frac{M_X}{m_{X}}-
 \nonumber \\
&-& \frac{b_{3,\Sigma_{8 H}}}{2\pi}\ln \frac{M_X}{m_{\Sigma_{8 H}}}
- \frac{b_{3,\Sigma_{8}}}{2\pi}\ln \frac{M_X}{m_{\Sigma_{8}}},
\end{eqnarray}
where ${\alpha}_i^{-1}(M_X)$ is a coupling constant evaluated from low-energy using the two-loop RGEs. The coefficients of the one-loop beta-functions are given by
\begin{eqnarray}
&&b_{1,{H_C}}=1/15, \ \ b_{1,X}=5/6, \nonumber \\
&& b_{2,X}=1/2,  \ \ b_{2,\Sigma_{3H}}=1/3, \nonumber \\
&& b_{3,{H_C}}=1/6, \ \ b_{3,X}=1/3,  \ \
b_{3, \Sigma_8}=1/2, \ \ b_{3, \Sigma_{8 H}}=1/2. 
\end{eqnarray}
We require
\begin{eqnarray}
\tilde{\alpha}_1^{-1}(M_X)= \tilde{\alpha}_2^{-1}(M_X) = \tilde{\alpha}_3^{-1}(M_X).
\end{eqnarray}
Then, the GUT scale particles masses can be determined. To simplify our expressions, we define 
\begin{eqnarray}
k_{X} &=& \frac{1}{2\pi} \ln \frac{M_X}{m_{X}}, \ \ 
k_{H_C} = \frac{1}{2\pi} \ln \frac{M_X}{m_{H_C}},
\nonumber \\
k_{8} &=& \frac{1}{2\pi} \ln \frac{M_X}{m_{\Sigma_8}}, \ \
k_{8H} = \frac{1}{2\pi} \ln \frac{M_X}{m_{\Sigma_{8H}}}~.
\end{eqnarray}
The conditions for gauge coupling unification are then written as 
\begin{eqnarray}
k_{8H} &=& \frac{1}{2}\left[-5 {\alpha}_1^{-1}(M_X) + 3{\alpha}_2^{-1}(M_X)+2 {\alpha}_3^{-1}(M_X)\right]
- k_{8}/2 + k_{X} +\frac{1}{4\pi} \ln 2 , \nonumber \\
k_{H_C} &=& \frac{5}{2} \left[ {\alpha}_1^{-1}(M_X) -3  {\alpha}_2^{-1}(M_X)+2{\alpha}_3^{-1}(M_X) - k_8  \right] 
- \frac{5}{4\pi} \ln 2.\label{eq:gut_corr_real} 
\end{eqnarray}
Since ${\alpha}_1^{-1}(M_X) -3  {\alpha}_2^{-1}(M_X)+2{\alpha}_3^{-1}(M_X) \sim -(4$-$5)$, it is difficult to achieve unification using only threshold corrections from fields of the minimal matter content.

\bibliographystyle{utphys}
\bibliography{refs}

\end{document}